\documentclass[twocolumn,showpacs,preprintnumbers,amsmath,amssymb]{revtex4}

\usepackage{graphicx}
\usepackage{dcolumn}
\usepackage{bm}

\usepackage{hyperref}
\usepackage{graphicx}
\usepackage{amsfonts}
\usepackage{amssymb}
\usepackage{amsmath}

\def\jqe{ IEEE J.\ Quantum Electron.\ }

\def\josa{ J.\ Opt.\ Soc.\ Am.\ }

\def\josab{ J.\ Opt.\ Soc.\ Am.\ B }

\def\oc{ Opt.\ Commun.\ }
\def\ol{ Opt.\ Lett.\ }

\def\pre{ Phys.\ Rev.\ E }
\def\prl{ Phys.\ Rev.\ Lett.\ }

\begin{document}

\title{Mode interaction in multi-mode optical fibers with Kerr effect}

\author{Sergey Leble}
 \email{leble@mif.pg.gda.pl}
\author{Bartosz Reichel}%
 \email{reichel@mif.pg.gda.pl}
\affiliation{%
Technical University of Gda\'{n}sk,\\
 ul. G.Narutowicza, 11  80-952, Gda\'{n}sk-Wrzeszcz, Poland
}%

\date{\today}

\begin{abstract}
We generalize the projection to orthogonal function basis (including
polarization modes) method for nonlinear (Kerr medium) fiber and use
this method in a case of two-mode waveguide. We consider orthogonal
Bessel functions basis that fit the choice of cylindrical geometry
of a fiber.  The coupled nonlinear Schr\"{o}dinger equations (CNLS)
are derived. Analytic expressions and numerical results for coupling
coefficients are given; the fiber parameters dependence is
illustrated by plots.
\end{abstract}

\maketitle

\section{introduction}

It is well known that Nonlinear Shr\"{o}dinger (NS) equation,
similar to celebrated KdV, presents the most universal physical
embedding of soliton behavior. However, its  experimental
realizations are still rather poor \cite{Rem} and rather demonstrate
the solitons than introduce a true physical tool for nonlinear
phenomena investigations. The main reason for this is its
uni-dimensionality and uni-directionality  - the embedding needs
some kind of projecting with exact formal and experimental
specification.

The important example of the theory is the optical guide (fiber)
propagation that isolates modes due to boundary conditions in a
fiber cross-section and the direction of propagation is fixed  by a
way in which the waveguide is excited. So the model based on NS is
essentially one dimensional, hence the fiber propagation is perhaps,
the best one for physical realization of soliton behavior of
electromagnetic wave-train. Nevertheless  the existing theory adopts
\cite{Mollenauer} the model of waveguide propagation, that looks
rather ''technical'' than physical one. The authors  use the
parameter of NS equation as empirical ones without explicit
expressions dependent on a whole geometry and physics of a
waveguide.

In the papers \cite{refMenyuk} and \cite{SamirGarth} authors
investigate polarization behavior of light in a Kerr fiber;
 additionally  the authors of \cite{SamirGarth}
took into account power exchange between modes. The authors of these
articles declare the complete basis expansion for the problem
solution but in fact use simplifications choosing for the basis
functions (for higher order modes) the Gaussian functions. This
approximation is used  for nonlinear coefficients (in our notation
$\mathbb{O}_{1}$) evaluation. The formulas  for the coupling
constants ($c_{ij}$ in \cite{GarthPask}) exhibit the fiber radius
dependence (proportional to $1/r_0^2$) that essentially differs from
ours.

By the work \cite{MonomodeCase} we started the detailed
investigations; we described and discussed a method to derive CNLS
(Coupled Nonlinear Schr\"{o}dinger) equations for a multi-mode
fibers  focusing on a mono-mode case (which leads to NS equations)
and made numerical calculations for it. We elaborated an approach
which do not apply potentials, each component of electromagnetic
field in cylindrical coordinates is represented directly as a series
in correspondent Bessel functions and exponentials
$J_{l}(\alpha_{ln}r) e^{il\varphi}$. Such scheme introduces natural
mode notion in a fiber cross-section which have obvious link with
the standard one (TE,TM modes).

Here we derive CNLS equations for two modes (where $l=0,\pm 1$ and
$n=0$ in both cases). Information about the parameters allows to
derive the explicit dependence of soliton properties on geometry and
material constant of a medium. We use the model of step index
waveguide with different refraction indexes (without assumption of
weakly guided fiber like in \cite{SamirGarth, GarthPask,refSnyder}).
The main result of this part of our study is the analytical
expression for mode interaction coefficients that exhibit rather
strong dependencies on waveguide radius.

 We take the fundamental mode (which is known
from linear a theory as HE11) considering it and the second mode
(TE01) as vectors of our basis. Both modes have different group
velocities, this difference influence on the coupling coefficients
magnitude. This model could be used in other waveguides with
different geometries (e.g. the elliptical one \cite{GarthPask}).
Both single-mode fibers with two polarizations and multi-mode fibers
are used in optical-switching devices \cite{KerrSw}.

Considering two modes with different propagation constants (this
mean $\delta\neq 0$), we expected that interaction between modes
occurs only at the beginning of propagation (we inquire the
propagation to begin from  excitation at one of waveguide's ends).
This can appear in the strong birefringent photonic crystal fibers
(PCF). In a consequence this interaction can lead to change power
intensity of modes. For example, such properties are very important
in case of photonic crystal fibers (PCF) or dispersion managed
solitons (DM).

It is necessary to point out that in the  cylindrical waveguide
even three modes can interact: HE11, TE01 and TM01. Modes TE01 and
TM01 have the same cut-off frequency and propagation constant
\cite{agrawal:book:NonFibOpt}. This case is very similar to single
mode with two polarizations case and we considered this in
\cite{MonomodeCase}. In case of HE11 - TE01 mode interaction, it
is needed to take into account two mode interaction with different
propagation constants. In this work we investigate only this
(HE11-TE01) interactions to simplify equations.

We start in the second section  from basic equations of linear and
some of nonlinear theory of electromagnetic fields. In third section
analytical solution of a given problem is presented. In fourth
section some of numerical calculations for nonlinear coefficients as
a function of the guide radius are given. The last section is a
comparison with other cases of analyzed equations and general
conclusion is presented.

\section{Basic equations}

Consider the propagation of optical pulse at isotropic medium and
choose a cylindrical dielectric waveguide with small Kerr
nonlinearity. The electric field can be written as
\begin{equation}
 E_{i}=\frac{1}{2}A_{i}e^{i\omega t}+c.c.,
\end{equation}
where $i=x,y,z$. We Introduce a linearly polarized field as
\cite{agrawal:book:NonFibOpt,Porsezian:OpticalSolitons}
\begin{equation}\label{EQ:BirifAxis}
E_{x,y}=\frac{1}{2}E_{x,y}^{+}+\frac{1}{2}E_{x,y}^{-},
\end{equation}
and the polarization vector component as
\begin{align}\nonumber
P_{z}&=\frac{3}{8}\chi_{xxxx}\varepsilon_{0}\left[|A_{z}|^{2}+\frac{2}{3}\left(|A_{x}|^{2}+|A_{y}|^{2}\right)\right]A_{z}e^{i\omega
t}\\
&+\frac{1}{3}\overline{A}_{z}\left(A_{x}^{2}+A_{y}^{2}\right)e^{i\omega
t}+ c.c.,
\end{align}
where $\overline{A}_{z}$ is the complex conjugate of $A_{z}$.

The electric field component in Bessel function basis, which is
standard for cylindrical waveguide can be written as
\cite{Shest,Hasegawa:book:solitons})
\begin{subequations} \label{E:EFieldCoeff_AiF_Kart}
\begin{align}\nonumber
E_{x}^{\pm}&(x,y,z,t)=\\
&\mp\frac{1}{2}\sum_{l,n}\frac{1}{\alpha_{ln}}
\partial_{z}\mathcal{A}^{\pm}_{ln}J_{l\pm 1}(\alpha_{ln}r)
e^{i(l\pm 1)\varphi}e^{i\omega t-ikz}+c.c.,
\end{align}
\begin{align}\nonumber
E_{y}^{\pm}&(x,y,z,t)=\\
&\frac{1}{2}\sum_{l,n}\frac{i}{\alpha_{ln}}\partial_{z}\mathcal{A}^{\pm}_{ln}J_{l\pm
1}(\alpha_{ln}r)e^{i(l\pm 1) \varphi}e^{i\omega t-ikz}+c.c.,
\end{align}
\end{subequations}
here $\alpha_{ln}$ is eigenvalue for linear cylindrical waveguide
problem
\begin{subequations} \label{E:BounduaryKvec}
\begin{eqnarray}
\alpha^{2}&=&\omega^{2}\varepsilon_{0}\mu_{0}\varepsilon_{1}-k^{2}\text{,\qquad $r\leq r_{0}$},\\
\beta^{2}&=&k^{2}-\omega^{2}\varepsilon_{0}\mu_{0}\varepsilon_{2}
\text{,\qquad $r>r_{0}$},
\end{eqnarray}
\end{subequations}
where $r_{0}$ is waveguide radius and the variable amplitude
$\mathcal{A}$ depends on propagation axis and time.

Using the multi-mode model \cite{MonomodeCase} (where the modes
orthogonality over the fiber cross section is used) we obtain
\begin{equation}\label{EQ:Equ_z_nonlin01}
\left(\square_{z}+\alpha^{2}_{01}\right)\mathcal{A}^{p}_{01}=\frac{2\varepsilon_{0}\mu_{0}}{\pi
N_{01}}\int\limits_{0}^{r_{0}}\int\limits_{0}^{2\pi}
rJ_{0}(\alpha_{01} r)\frac{\partial^2}{\partial t^2}P_{z}d\varphi
dr,
\end{equation}
\begin{equation}\label{EQ:Equ_z_nonlin11}
\left(\square_{z}+\alpha^{2}_{11}\right)\mathcal{A}^{p}_{11}=\frac{2\varepsilon_{0}\mu_{0}}{\pi
N_{11}}\int\limits_{0}^{r_{0}}\int\limits_{0}^{2\pi}
rJ_{1}(\alpha_{11} r)e^{-i\varphi}\frac{\partial^2}{\partial
t^2}P_{z}d\varphi dr,
\end{equation}
where $\square_{z}$ is defined by
\begin{equation}\label{EQ:operator1z}
\square_{z} =
\mu_{0}\varepsilon_{0}\varepsilon\frac{\partial^{2}}{\partial
t^{2}} -  \frac{\partial^{2}}{\partial z^{2}}.
\end{equation}
The  coefficient at the r.h.s. denominators has the following form
\begin{equation}\label{E:BesselOrhtoNormal}
N_{nl}=\frac{r_{0}^{2}}{2}\left[J_{l}^{2}(\alpha_{ln}
r_{0})-J_{l-1}(\alpha_{ln} r_{0})J_{l+1}(\alpha_{ln} r_{0})\right].
\end{equation}

In the case of  a single-mode fiber (with two polarization) we
could omit the difference between group velocities, assuming that
the fiber is isotropic and do not have bending
\cite{agrawal:book:NonFibOpt}. However in a multi-mode fiber it
has to be taken into account, because different modes have
substantially different group velocities. We describe it
introducing the co-ordinate system which moves with average group
velocity
\begin{subequations}\label{E:SlowCoo}
\begin{eqnarray}
  \xi &=& \sigma z, \\
  \tau &=& (t-\beta'z)\epsilon,\\
  \beta'&=&\frac{k_{01}'+k_{11}'}{2}.
\end{eqnarray}
\end{subequations}
Next step is introducing a slowly varying amplitude of the wave
envelope \cite{Hasegawa:book:solitons} in the form
\begin{equation}\label{E:SlowAmplitudeDef}
  \sigma X^{\pm}(\tau,\xi)e^{-ikz},
\end{equation}
and $X$ have the same unit as electrical field [V/m].
\section{Two modes interaction}

First, due to the isotropic material assumption, we have
$k_{01}^{+}=k_{01}^{-}$ and $k_{11}^{+}=k_{11}^{-}$. Plugging a
solution for electromagnetic field \cite{refMenyuk,
agrawal:book:NonFibOpt} in to the left hand side of the equations
\eqref{EQ:Equ_z_nonlin01}-\eqref{EQ:Equ_z_nonlin11} yields
 four equations (two modes and each mode have two polarizations):
\begin{subequations}\label{E:CNLSE}
\begin{multline}
i\partial_{\xi}X^{\pm}_{01}-
i\delta\partial_{\tau}X^{\pm}_{01}+\frac{\epsilon^{2}k''}{2\sigma}
\partial_{\tau\tau}X^{\pm}_{01}=\\
\mathbb{P}_{01}\left[\mathbb{O}_{1}|X_{01}^{\pm}|^{2}+\mathbb{O}_{2}|X_{01}^{\mp}|^{2}+\mathbb{O}_{3}|X_{11}^{\pm}|^{2}\right]X_{01}^{\pm}\\
+\mathbb{P}_{01}\left[\mathbb{O}_{4}|X_{11}^{\mp}|^{2}+\mathbb{O}_{5}X_{11}^{\mp}X_{11}^{\pm
*}+\mathbb{O}_{6}X_{11}^{\pm}X_{11}^{\mp *}\right]X_{01}^{\pm},
\end{multline}
\begin{multline}
i\partial_{\xi}X^{\pm}_{11}+i\delta\partial_{\tau}X^{\pm}_{11}+\frac{\epsilon^{2}k''}{2\sigma}
\partial_{\tau\tau}X^{\pm}_{11}=\\
\mathbb{P}_{11}\left[\mathbb{Q}_{1}|X_{11}^{\pm}|^{2}+\mathbb{Q}_{2}|X_{11}^{\mp}|^{2}+\mathbb{Q}_{3}|X_{01}^{\pm}|^{2}\right]X_{11}^{\pm}\\
+\mathbb{P}_{11}\left[\mathbb{Q}_{4}|X_{01}^{\mp}|^{2}+\mathbb{Q}_{5}X_{01}^{\mp}X_{01}^{\pm
*}+\mathbb{Q}_{6}X_{01}^{\pm}X_{01}^{\mp *}\right]X_{11}^{\pm},
\end{multline}
\end{subequations}
where
\begin{equation}
\delta=\frac{1}{2}\left(k'_{01}-k'_{11}\right).
\end{equation}

The coupling coefficients have the following form
\begin{equation}
\mathbb{O}_{1}= \int\limits_{0}^{r_{0}}
r\left[J_{0}^{4}(\alpha_{01}r)+\frac{4k_{01}^{2}}{3\alpha_{01}^{2}}J_{1}^{2}(\alpha_{01}r)J_{0}^{2}(\alpha_{01}r)\right]dr,
\end{equation}
\begin{align}\nonumber
\mathbb{Q}_{1}&= \int\limits_{0}^{r_{0}}
rJ_{1}^{4}(\alpha_{11}r)dr\\
&+\int\limits_{0}^{r_{0}} r\left[\frac{2k_{11}
^{2}}{3\alpha_{11}^{2}}J_{1}^{2}(\alpha_{11}r)\left(J_{0}^{2}(\alpha_{11}r)+J_{2}^{2}(\alpha_{11}r)\right)\right]dr,
\end{align}

\begin{equation}
\mathbb{O}_{2}= \int\limits_{0}^{r_{0}}
r2J_{0}^{4}(\alpha_{01}r)dr,
\end{equation}

\begin{align}\nonumber
\mathbb{Q}_{2}&= \int\limits_{0}^{r_{0}}
r2J_{1}^{4}(\alpha_{11}r)dr\\
&+\int\limits_{0}^{r_{0}} r\left[\frac{2k_{11}
^{2}}{3\alpha_{11}^{2}}J_{1}^{2}(\alpha_{11}r)\left(J_{0}(\alpha_{11}r)+J_{2}(\alpha_{11}r)\right)^{2}\right]dr,
\end{align}

\begin{align}\nonumber
\mathbb{O}_{3}&= \int\limits_{0}^{r_{0}}
r2J_{1}^{2}(\alpha_{11}r)J_{0}^{2}(\alpha_{01}r)dr\\%
&\nonumber+\int\limits_{0}^{r_{0}}
r\frac{2k_{11}^{2}}{3\alpha_{11}^{2}}J_{0}^{2}(\alpha_{01})\left(J_{2}^{2}(\alpha_{11}r)+
J_{0}^{2}(\alpha_{11}r)\right)%
dr%
\\\nonumber+&\int\limits_{0}^{r_{0}} r%
\frac{2k_{01}k_{11}}{3\alpha_{01}\alpha_{11}}J_{1}(\alpha_{01}r)J_{1}(\alpha_{11}r)J_{0}(\alpha_{01}r)J_{2}(\alpha_{11}r)%
dr\\-&\int\limits_{0}^{r_{0}} r%
\frac{2k_{01}k_{11}}{3\alpha_{01}\alpha_{11}}J_{1}(\alpha_{01}r)J_{1}(\alpha_{11}r)J_{0}(\alpha_{01}r)J_{0}(\alpha_{11}r)%
dr,
\end{align}

\begin{align}\nonumber
\mathbb{Q}_{3}&= \int\limits_{0}^{r_{0}}
r2J_{1}^{2}(\alpha_{11}r)J_{0}^{2}(\alpha_{01}r)dr\\\nonumber
&+\int\limits_{0}^{r_{0}}
r\frac{4k_{01}^{2}}{3\alpha_{01}^{2}}J_{1}^{2}(\alpha_{11}r)J_{1}^{2}(\alpha_{01}r)dr\\\nonumber
&+%
\int\limits_{0}^{r_{0}} r
\frac{2k_{01}k_{11}}{3\alpha_{01}\alpha_{11}}J_{1}(\alpha_{11}r)J_{1}
(\alpha_{01}r)J_{0}(\alpha_{01}r)J_{2}(\alpha_{11}r)dr,
\\&-%
\int\limits_{0}^{r_{0}} r
\frac{2k_{01}k_{11}}{3\alpha_{01}\alpha_{11}}J_{1}(\alpha_{11}r)J_{1}
(\alpha_{01}r)J_{0}(\alpha_{01}r)J_{0}(\alpha_{11}r)dr,
\end{align}

\begin{align}
\mathbb{O}_{4}&= \int\limits_{0}^{r_{0}}
r2J_{1}^{2}(\alpha_{11}r)J_{0}^{2}(\alpha_{01}r)dr+\\\nonumber
&+\int\limits_{0}^{r_{0}}
r\frac{2k_{11}^{2}}{3\alpha_{11}^{2}}J_{0}^{2}(\alpha_{01})\left(J_{2}^{2}(\alpha_{11}r)+J_{0}^{2}(\alpha_{11}r)\right)dr\\\nonumber%
&+\int\limits_{0}^{r_{0}}r%
\frac{2k_{01}k_{11}}{3\alpha_{01}\alpha_{11}}J_{1}(\alpha_{01}r)J_{1}(\alpha_{11}r)J_{0}(\alpha_{01}r)J_{0}(\alpha_{11}r)dr\\\nonumber%
&-\int\limits_{0}^{r_{0}}r
\frac{2k_{01}k_{11}}{3\alpha_{01}\alpha_{11}}J_{1}(\alpha_{01}r)J_{1}(\alpha_{11}r)J_{0}(\alpha_{01}r)J_{2}(\alpha_{11}r)dr,\nonumber%
\end{align}
\begin{align}
\mathbb{Q}_{4}&= \int\limits_{0}^{r_{0}}
r2J_{1}^{2}(\alpha_{11}r)J_{0}^{2}(\alpha_{01}r)\\\nonumber
&+\frac{4k_{01}^{2}}{3\alpha_{01}^{2}}J_{1}^{2}(\alpha_{11}r)J_{1}^{2}(\alpha_{01}r)dr\\\nonumber
&+\int\limits_{0}^{r_{0}} r
\frac{2k_{01}k_{11}}{3\alpha_{01}\alpha_{11}}J_{1}(\alpha_{11}r)J_{1}
(\alpha_{01}r)J_{0}(\alpha_{01}r)J_{0}(\alpha_{11}r)dr\\\nonumber
&+\int\limits_{0}^{r_{0}} r
\frac{2k_{01}k_{11}}{3\alpha_{01}\alpha_{11}}J_{1}(\alpha_{11}r)J_{1}
(\alpha_{01}r)J_{0}(\alpha_{01}r)J_{2}(\alpha_{11}r)dr,\\\nonumber
\end{align}

\begin{equation}
\mathbb{Q}_{5,6}=\mathbb{O}_{5,6}= \int\limits_{0}^{r_{0}}
r2J_{1}^{2}(\alpha_{11}r)J_{0}^{2}(\alpha_{01}r)dr,
\end{equation}
and
\begin{eqnarray}
\mathbb{P}_{l1}=\frac{3\omega^{2}\chi_{xxxx}}{32
N_{l1}k_{l1}c^{2}},
\end{eqnarray}
where $l=0,1$ and $P_{l1}$ have units $[(V^{2}m)^{-1}]$ while
$\chi$ is in $[m^{2}/V^{2}]$.

The coefficients $\mathbb{O}_{1}$ and $\mathbb{O}_{2}$ are the same
as coefficients for one mode fiber with $l=0$ and $n=1$. The
$\mathbb{Q}_{1}$ and $\mathbb{Q}_{2}$ coefficients for the
multi-mode fiber are the same as for one mode one with $l=1$ and
$n=0$. This part of equations describe interaction between
polarization mode with same number $l$ and have been analyzed in
\cite{MonomodeCase}.

Rest of the coefficients describe the coupling between two
different modes (with different $l$ number and theirs
polarizations).

Notice that coefficient $\mathbb{O}_{5}$, $\mathbb{O}_{6}$,
$\mathbb{Q}_{5}$ and $\mathbb{Q}_{6}$ have same values because they
describe mixed interaction (between different modes and different
polarization).

\section{Numerical results}
First numerically evaluate eigenvalues $\alpha_{01}$ and
$\alpha_{11}$
 from Hondros-Debye equation. In the next steep integrals with
Bessel function are calculated numerically. Eigenvalues are
evaluated without approximation of weakly guided fiber where we
have $\varepsilon_{1} \approx \varepsilon_{2}$ \cite{refSnyder}.

\begin{figure}[htbp]
\centering
\includegraphics[width=0.44\textwidth]{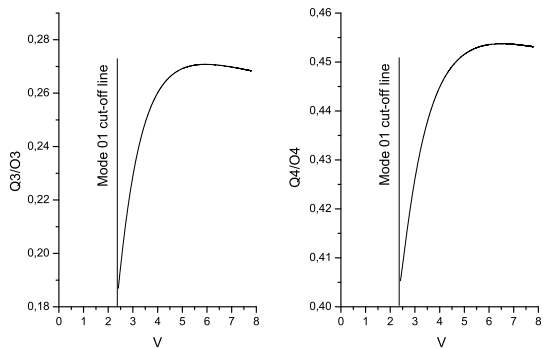}
\caption{Numerical results for coupling coefficient for mode $01$
and $11$} \label{PIC:Q3O3}
\end{figure}

\begin{figure}[htbp]
\centering
\includegraphics[width=0.4\textwidth]{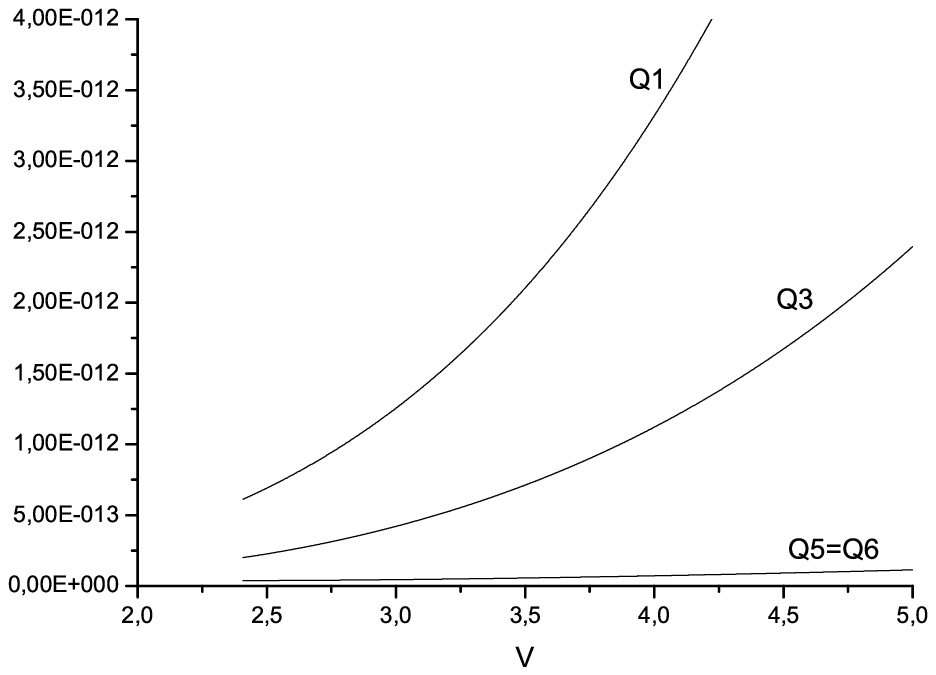}
\caption{Numerical results for coupling coefficient, difference
between $\mathbb{Q}_{5}$, $\mathbb{Q}_{6}$ and other coefficient.
} \label{PIC:Small56}
\end{figure}

The results for inter-mode influence are shown at the figure
\ref{PIC:Q3O3}. Note that the mode $01$ have cutoff frequency near
$V\approx 2.4$ for that reason the coefficients have different
behaviour.
Coefficients $\mathbb{O}_{5}$, $\mathbb{O}_{6}$, $\mathbb{Q}_{5}$
and $\mathbb{Q}_{6}$ describe mixed interactions (different modes
and different polarizations) and for isotropic medium have same
value. Figure \ref{PIC:Small56} shows difference between
coefficients
($\mathbb{O}_{5}=\mathbb{O}_{6}=\mathbb{Q}_{5}=\mathbb{Q}_{6}$ is
smallest and have smallest increment). All of this coefficients are
smallest than $\mathbb{O}_{1} - \mathbb{O}_{4}$ coefficients.

In our calculation we defined normalized frequency as
\begin{equation}
V=\frac{\omega}{c}r_{0}\sqrt{\varepsilon_1-\varepsilon_2},
\end{equation}
and in numerical calculations we used physical parameters with
following values:
\begin{subequations}\label{NumParam}
\begin{eqnarray}
    \omega &=& 12.2*10^{14} \quad \text{Hz}\quad (\lambda\approx 1.54\mathrm{\mu m} ),\\
    \varepsilon_1&=&2.25 \quad\text{(ref. index $1.5$)},\\
    \varepsilon_2&=&1.96 \quad\text{(ref. index $1.4$)},\\
    r_0&& \text{from $1.2*10^{-6}$m to $10*10^{-6}$m}.
\end{eqnarray}
\end{subequations}

\section{Conclusion}
\begin{figure}[htbp]
\centering
\includegraphics[width=0.4\textwidth]{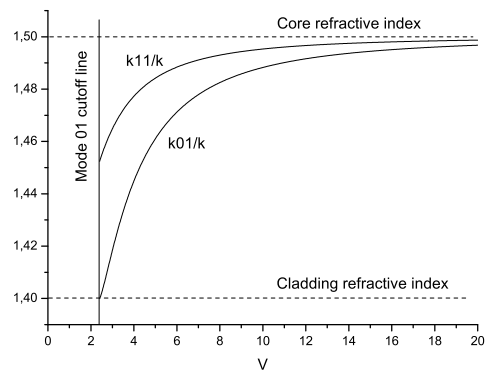}
\caption{Difference in propagation constants between the 11 mode
and the 01 mode, where $k=\omega/c$} \label{propagationConst}
\end{figure}

In this paper we considered the influence of Kerr nonlinearity on
the mode coupling in the case of two-mode fiber. The results show
how the modes influence each other and we expect higher influence
when $V$ is bigger. The reason is that the propagation constants
$k_{01}$ for higher $V$ is roughly $k_{11}$ (fig.
\ref{propagationConst}).

We can reconsider equation \eqref{E:CNLSE} again but with
simplification, that we do not take into the account polarization
interaction. Choosing two modes with the same polarization yield:
\begin{subequations}\label{E:CNLSE_SigPol}
\begin{multline}
i\partial_{\xi}X_{01}-
i\delta\partial_{\tau}X_{01}+\frac{\epsilon^{2}k''}{2\sigma}
\partial_{\tau\tau}X_{01}=\\
\mathbb{P}_{01}\left[\mathbb{O}_{1}|X_{01}|^{2}+\mathbb{O}_{3}|X_{11}|^{2}\right]X_{01},
\end{multline}
\begin{multline}
i\partial_{\xi}X_{11}+i\delta\partial_{\tau}X_{11}+\frac{\epsilon^{2}k''}{2\sigma}
\partial_{\tau\tau}X_{11}=\\
\mathbb{P}_{11}\left[\mathbb{Q}_{1}|X_{11}|^{2}+\mathbb{Q}_{3}|X_{01}|^{2}\right]X_{11}.
\end{multline}
\end{subequations}

In this case we have only four parameters (two for self phase
modulation and two for cross phase modulation), which describe only
interaction between modes. In boundary when we have two modes with
the same group velocity ($\delta=0$), we get system of equations
like this in case of polarization interaction \cite{MonomodeCase}.

Method which we used for a cylindrical waveguide  is easily
reformulated for a  different waveguide shape where second-modes
(and higher modes) are important, for example in the elliptical
waveguide \cite{GarthPask}. This method also can be used for a
photonic crystal fiber, directly or within the approximations of the
approach of \cite{DD05}. Here we have delivered all calculations for
 isotropic medium but there is a possibility to make it for
anisotropic medium.

\begin{acknowledgments}
 The work is supported by the Polish Ministry of
Scientific Research and information Technology grant
PBZ-Min-008/P03/2003.
\end{acknowledgments}



\begin{thebibliography}{99}
\bibitem{Rem}M. Remoissenet Waves Called Solitons: Concepts and Experiments
Springer 1996.

\bibitem{Mollenauer}
L.F.Mollenauer, R.H.Stolen, J.P.Gordon, \prl, {\bf 45,} 1095
(1980).

\bibitem{refMenyuk} C.R.Menyuk, \jqe, {\bf QE-23(2),} 174 (1987).

\bibitem{SamirGarth}W.Samir, S.J.Garth,  \oc. {\bf 94,} 373 (1992).

\bibitem{GarthPask}S.J.Garth, C.Pask, \josab, {\bf  9,} 243
(1992).


\bibitem{MonomodeCase} S.B.Leble, B.Reichel, \emph{Polarization mode interaction equations in optical fibers with Kerr
effect}, arxiv.org:physics/0411255, 2004, (Submitted to Physica
D).
\bibitem{refSnyder} A.W.Snyder, W.R. Young, \josa, {bf 3,}  297
(1978)



\bibitem{KerrSw}H.G.Park, C.C.Pohalski, B.Y.Kim, \ol, {\bf
13,} 776 (1988).

\bibitem{agrawal:book:NonFibOpt} G.P.Agrawal, {\em Nonlinear fiber optics}, Academic Press, 1997.

\bibitem{Porsezian:OpticalSolitons} V.C.Kuriakose, K.Porsezian, {\em Optical solitons. Theoretical and experimental challenges}, Springer, 2002.

\bibitem{Hasegawa:book:solitons} Y.Kodama, A.Hasegawa, {\em Solitons in optical communication}, Clarendon press, Oxford, 1995.


\bibitem{Shest}H.W.Sch\"{u}rmann, Y.Smirnov, Y.Shestopalov, \pre,
{\bf 71,} 016614 (2005).


\bibitem{DD05} B.Reichel, S.B.Leble, {\em Projection to orthogonal function basis method
for nonlinear multi-mode fiber}, Conference material - Day on
Diffraction St.Petersburg 2005.


\end{thebibliography}
\end{document}